\documentclass[reprint,prx,aps,twocolumn,showpacs,reprint,superscriptaddress,nofootinbib]{revtex4-1}
\usepackage{amsmath,amssymb,hyperref}
\usepackage{graphicx}
\usepackage{datetime}
\usepackage{braket}
\usepackage{xfrac}
\usepackage{upgreek}
\usepackage[usenames,dvipsnames]{color}
\usepackage{natbib}
\usepackage{hyperref}
\usepackage{appendix}
\usepackage{titlesec}
\usepackage{ulem}

\makeatletter
\AtBeginDocument{\let\LS@rot\@undefined}
\makeatother

\definecolor{MyGreen}{rgb}{0,0.4,0}

\newcommand{\gTwo}{\ensuremath{g^{(2)}}}

\newcommand{\nmol}{\ensuremath{N_{\text{mol}}}}

\newcommand{\nex}{\ensuremath{N_{\text{ex}}}}

\newcommand{\Imperial}{
Physics Department, Blackett Laboratory, Imperial College London, Prince Consort Road, SW7 2AZ, UK}
\newcommand{\ImperialCQD}{Centre for Doctoral Training in Controlled Quantum Dynamics, Imperial College London, Prince Consort Road, SW7 2AZ, UK}
\begin{document}
\title{Non-stationary Statistics and Formation Jitter in Transient Photon Condensation}
%
%
%
\author{Benjamin T. Walker} \affiliation{\Imperial}\affiliation{\ImperialCQD}
\author{Jo\~ao D. Rodrigues}\email[Correspondence to ]{j.marques-rodrigues@imperial.ac.uk}\affiliation{\Imperial}
\author{Himadri S. Dhar}\affiliation{\Imperial}
\author{Rupert F. Oulton}\affiliation{\Imperial}
\author{Florian Mintert}\affiliation{\Imperial}
\author{Robert A. Nyman}\affiliation{\Imperial}

%
\date{\today, \ampmtime}
%
\begin{abstract}
While equilibrium phase transitions are well described by a free-energy landscape, there are few tools to describe general features of their non-equilibrium counterparts. On the other hand, near-equilibrium free-energies are easily accessible but their full geometry is only explored in non-equilibrium conditions, e.g. after a quench. In the particular case of a non-stationary system, however, the concepts of an order parameter and free energy become ill-defined, and a comprehensive understanding of non-stationary (transient) phase transitions is still lacking. Here, we probe transient non-equilibrium dynamics of an optically pumped, dye-filled microcavity which exhibits near-equilibrium Bose-Einstein condensation under steady-state conditions. By rapidly exciting a large number of dye molecules, we quench the system to a far-from-equilibrium state and, close to a critical excitation energy, find delayed condensation, interpreted as a transient equivalent of critical slowing down. We introduce the two-time, non-stationary, second-order correlation function, $g^{(2)} (t_1, t_2)$, as a powerful experimental tool for probing the statistical properties of the transient relaxation dynamics. In addition to number fluctuations near the critical excitation energy, we show that transient phase transitions exhibit a different form of diverging fluctuations, namely timing jitter in the growth of the order parameter. This jitter is seeded by the randomness associated with spontaneous emission, with its effect being amplified near the critical point. The experimental results are accurately described by a microscopic model of light-matter interactions. The general character of our observations is then discussed based on the geometry of effective free-energy landscapes. We thus identify universal features, such as the formation timing jitter, for a larger set of systems undergoing transient phase transitions. Our results carry immediate implications to diverse systems, including micro- and nano-lasers and growth of colloidal nanoparticles.
\end{abstract}
%
\maketitle
%
%
%
\section{Introduction}
\par
As far back as 1873~\cite{gibbs1873method}, Gibbs appreciated the power of a geometric treatment of free-energies for understanding thermodynamic equilibria. Later, Jaynes generalised these ideas to non-equilibrium systems~\cite{jaynes1986predictive} and described how in cases where an entropy surface can be defined, its geometry fully determines how a ``bubble'' of probability perturbed by fluctuations evolves in time. Such a probability bubble encodes the statistical properties of a non-equilibrium order parameter. A key prediction is that, in regions where the entropy surface is concave, or equivalently, the free-energy is convex, the probability bubble becomes unstable, leading to a bifurcation. This instability has tremendous implications for the dynamics of order parameters in non-equilibrium systems, particularly in cases where a steady state has not or cannot be reached.
\par
The evolution of an ordered phase driven through configuration space by fluctuations (thermal, or quantum) is well described by an order parameter which evolves through a free-energy landscape~\cite{jaynes1957information, chipot2007free}, even though not all the free-energy surface is explored. Near a steady state, for instance, fluctuations around thermal equilibrium allow the free-energy landscape to be locally probed. An equivalent picture can be taken in non-equilibrium systems that can be mapped onto equilibrium statistical mechanics. A canonical example is the laser \cite{degiorgio1970}, a fundamentally non-equilibrium system whose steady-state can be described as the minimum of a properly defined effective free-energy, corresponding to a detailed balance between driving and dissipation. 
\par
While the previous arguments are relevant for systems close to a steady state, a sudden parameter change, often called a quench, necessarily brings the system sufficiently far-from-equilibrium to question the validity of the such approaches. The meaning of a quench depends on context and, in particular, one can distinguish between Hamiltonian and non-Hamiltonian cases. The former consist of time-dependent variations in some sort of interaction term, involved, for instance, in the Mott insulator-superfluid transition~\cite{Greiner2002, Chen2011} or the build-up of anti-ferromagnetic correlations in Ising models~\cite{Guardado2018}. Non-Hamiltonian quenches contain a more general class of processes. In cold atoms, for instance, the Kibble-Zurek mechanism~\cite{Navon2015, Weiler2008} is observed by evaporatively cooling the system at a finite rate, quenching the system through a BEC phase transition. We shall refer to a quench as a sudden change in one of the system parameters that brings it to a far-from-equilibrium state, without affecting its Hamiltonian. 
\par
Here, we study the transient dynamics of photon condensation that follows a quench in a dye-filled optical microcavity. Besides measuring the ensemble-averaged photon number dynamics, we introduce the non-stationary, two-time, second-order correlation function $g^{(2)}(t_1,t_2)$. It provides access to the statistical properties of the photon condensation transition, and is particularly relevant in non-stationary systems, when the full knowledge of individual realizations is inaccessible. While the usual stationary correlation function, \gTwo($\tau$), accurately accounts for fluctuations in steady-state, $g^{(2)}(t_1,t_2)$ is the appropriate quantity to describe the evolution of transient, non-equilibrium systems. The averaged condensate intensity as a function of time shows width broadening, a manifestation of diverging jitter in the condensate formation time upon approaching the critical excitation energy. This effect is directly witnessed by distinctive off-diagonal anti-correlations in $g^{(2)}(t_1,t_2)$ and originates from quantum fluctuations associated with spontaneous emission. By properly defining an effective free-energy, we argue that such jitter is a universal feature of transient phase transitions in systems obeying relatively general conditions on the convexity of their free-energy landscape.
\par
This paper is organized as follows: In Sec.~(\ref{experiment}) we present a microscopic cavity model, the experimental setup, and discuss the experimental results. By averaging over all forms of correlations and fluctuations we demonstrate a transient equivalent of critical slowing down. Retaining two-time correlations, we reveal the jitter in the condensate formation time. From the theoretical model, we construct a quantum trajectories simulation and a quantum regression approach. The former, by keeping correlations to all orders, better describes the results. In Sec.~(\ref{FE}), we explicitly convert the microscopic model into an effective free-energy landscape and present an approach that captures the qualitative features of the system dynamics, including formation jitter. The analysis in terms of the geometry of the free-energy landscape, being independent of the microscopic details of our particular system, generalizes the phenomenology observed in here to a broader class of systems. In particular, we discuss immediate implications in micro- and nano-lasers and in the growth of colloidal nanoparticles.
%
%
\section{Non-stationary Statistics}
\label{experiment}
%
%
\subsection{Microscopic Cavity Model}
\label{model}
\par
Despite the fundamentally multi-mode character of our optical cavity, the phenomenology described here is essentially that of a single-mode system. Cavity excitations (photons and excited molecules) can be lost by two processes: mirror transmission and molecular spontaneous emission into free-space, at rates $\kappa$ and $\Gamma_\downarrow$, respectively.  The essentials of the cavity dynamics are described by the density operator $\rho$, for both photons and molecules, which obeys the master equation~\cite{Kirton2013, kirton2015, Ozturk2019}
\begin{eqnarray}
\frac{d\rho}{dt} &=& -i[{H_0},\rho] + \kappa \mathcal{L}[\hat{a}]\rho + \sum_{k=1}^{\nmol} \left\{\Gamma_\downarrow ~\mathcal{L}[\sigma_k^{-}] + \Gamma_\uparrow~ \mathcal{L}[\sigma_k^{+}] \right.\nonumber\\
&+&  \left.A~ \mathcal{L}[\hat{a}~\sigma_k^{+}] +E~ \mathcal{L}[\hat{a}^\dag\sigma_k^{-}]\right\}\rho,
\label{M_eq}
\end{eqnarray}
with the Hamiltonian for the bare cavity $H_0 = \hat{a}^\dag \hat{a}$, $E$ and $A$ the cavity photon emission and absorption rates, respectively, $\Gamma_\uparrow$ the incoherent (external) pumping rate and \nmol\ the total number of molecules inside the cavity. Due to the high collision rate between dye and solvent molecules, all the relevant cavity processes, including light-matter interactions, are incoherent.
\par
Mean-field rate equations are obtained by taking expectation values and neglecting correlation terms in Eq.~(\ref{M_eq}). The number of cavity photons $n = \langle\hat{a}^\dag \hat{a} \rangle$ and fraction of excited molecules $f = \sum_k \langle\sigma^+_k\sigma^-_k\rangle / \nmol$, with $\langle \cdot \rangle$ denoting the (quantum-mechanical) ensemble average, are then determined by
\begin{equation}
\dot n = (E+A)(f-f_c)n+Ef, \quad \text{and}
\label{ndot}
\end{equation}
\begin{equation}
\label{fdot}
\dot f = -\Gamma_{\downarrow} f + \frac{An}{\nmol}(1-f) - \frac{E(n+1)}{\nmol}f + \Gamma_{\uparrow}(1-f).
\end{equation}
Here, the critical excitation fraction is defined as
\begin{equation}
f_c = \frac{\kappa + A}{A + E}
\end{equation}
and, in the limit of high number of photons, corresponds to the transition point between a net increase or decrease in the number of cavity photons with time. At $f=f_c$, and in the absence of pumping and losses ($\kappa = \Gamma_{\downarrow} = \Gamma_{\uparrow} = 0$), an equilibrium between molecular excitations and photons is established by a principle of detailed balance~\cite{schmitt2015thermalization}. The photon number in this equilibrium state would show a phase transition as the total number of cavity excitations, $N_\text{ex} = n + f N_\text{mol}$, which is the control parameter, is increased. The photon number, or order parameter, ranges from a disordered phase ($n \lesssim 1$) dominated by spontaneous emission to an ordered phase ($n \gg 1$) dominated by stimulated emission. By exciting a large number of dye molecules over a short period of time, the cavity can be quenched through this phase transition to a far-from-equilibrium state. The subsequent relaxation dynamics correspond to a non-stationary, {\it transient} counterpart of the equilibrium phase transition described above. However, the non-linear coupling between photons and molecular excitations occurring during this transient relaxation process gives rise to non-trivial fluctuation and correlation properties. Finally, given its lossy character, the light will transition back to the phase dominated by spontaneous emission before all excitations are lost.
\par
A few notes are in order regarding the multi-mode nature of our cavity. Within the single-mode approximation, the rate term $\Gamma_\downarrow$ accounts for emission both into free-space and cavity modes that do not reach the regime of stimulated emission (do not condense), which will be discussed in more detail in Sec.~(\ref{ensemble_av}). Also, and despite not being relevant for the results discussed here, effects associated with the multi-mode character as well as spatially resolved molecular reservoirs have been appreciated in the context of gain clamping~\cite{keeling2016} and decondensation mechanisms~\cite{Hesten2018}.
%
%
\subsection{Experimental Set-up}
\label{setup}
\begin{figure}
\centering
\includegraphics[width = 0.45\textwidth]{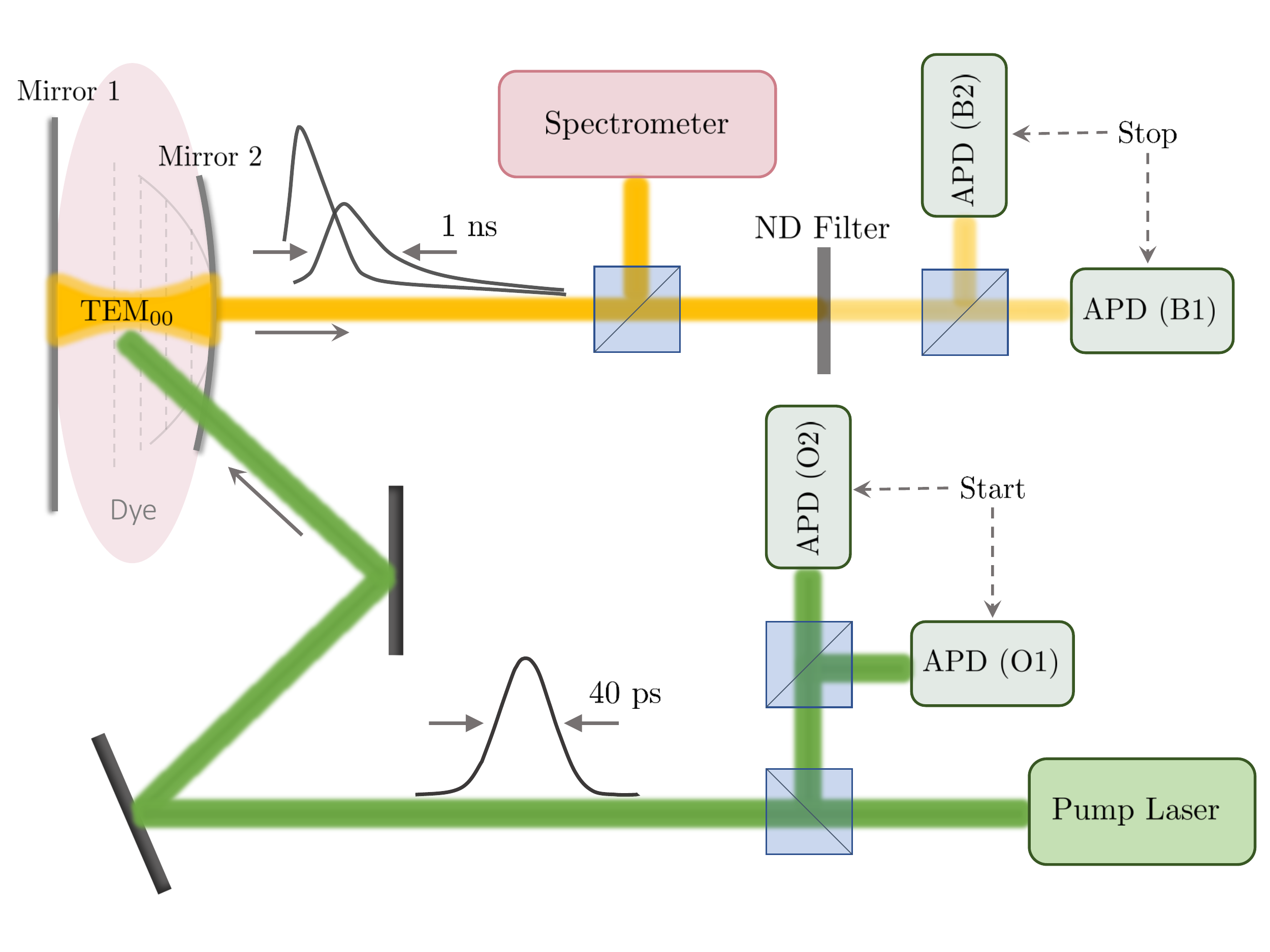}
\caption{Schematic representation of the experimental set-up. The dye-filled microcavity is composed of one planar (1) and one spherical (2) mirror.  The pulsed pump (green) is transmitted through the cavity at an angle of approximately 52$^{\circ}$. The transverse ground-state mode (yellow) leaks through one of the mirrors and is directed both into a spectrometer and a pair of single-photon detectors, B1 and B2, in an Hanbury-Brown-Twiss arrangement. A second pair of detectors, O1 and O2, is used to time the beginning of the experiment.}
\label{schematic}
\end{figure}
\par
The experimental configuration is sketched in Fig.~(\ref{schematic}). The optical cavity is composed of one planar and one spherical mirror of 0.25 meters radius of curvature, which traps the photons. The cavity is filled with a 2 mM solution of Rhodamine-6G in ethylene glycol. All the essential dynamics occur at the 10th longitudinal mode, corresponding to a cavity length of approximately 2~$\mu$m. A 40~ps laser pulse at 532 nm, typically ranging from 0.5 to 2 nJ in energy, is used to rapidly excite the molecules, quenching the cavity to a far-from-equilibrium state. In response, a much longer pulse ($\gtrsim$~1~ns) of light leaks from the cavity mirrors, the exact temporal shape of which depends both on the cavity parameters (loss rate, dye concentration, emission and absorption rates) as well as the number of molecular excitations that follow the pump pulse, which is the control parameter used to select the different dynamical phases. A portion of pump light is directed onto two saturated avalanche single-photon detectors (APDs), O1 and O2, where a coincident detection is used as a time stamp for the beginning of the experiment, with a measured uncertainty of about 10 ps. The cavity output light is directed onto two unsaturated APDs, B1 and B2, with an average of 0.1 detections per pulse, on each detector. The experiment is conducted at a repetition rate of 11 kHz. Such a low repetition rate ensures a complete decay of all excitations and statistical independence between different realizations. We describe the experimental results in the form of three sets:
\begin{itemize}
\item[--]\textit{zero-time} statistics: full time-averaged cavity output;
\item[--]\textit{one-time} statistics: time-resolved, but averaged over all forms of fluctuations and correlations in the cavity output;
\item[--]\textit{two-time} statistics: unequal time, cross-correlated signal from detectors B1 and B2, providing access to fluctuations in the cavity output.
\end{itemize}
%
%
\subsection{Zero-time Statistics}
\label{threshold}
\begin{figure}[t]
\centering
\includegraphics[width = 0.5\textwidth]{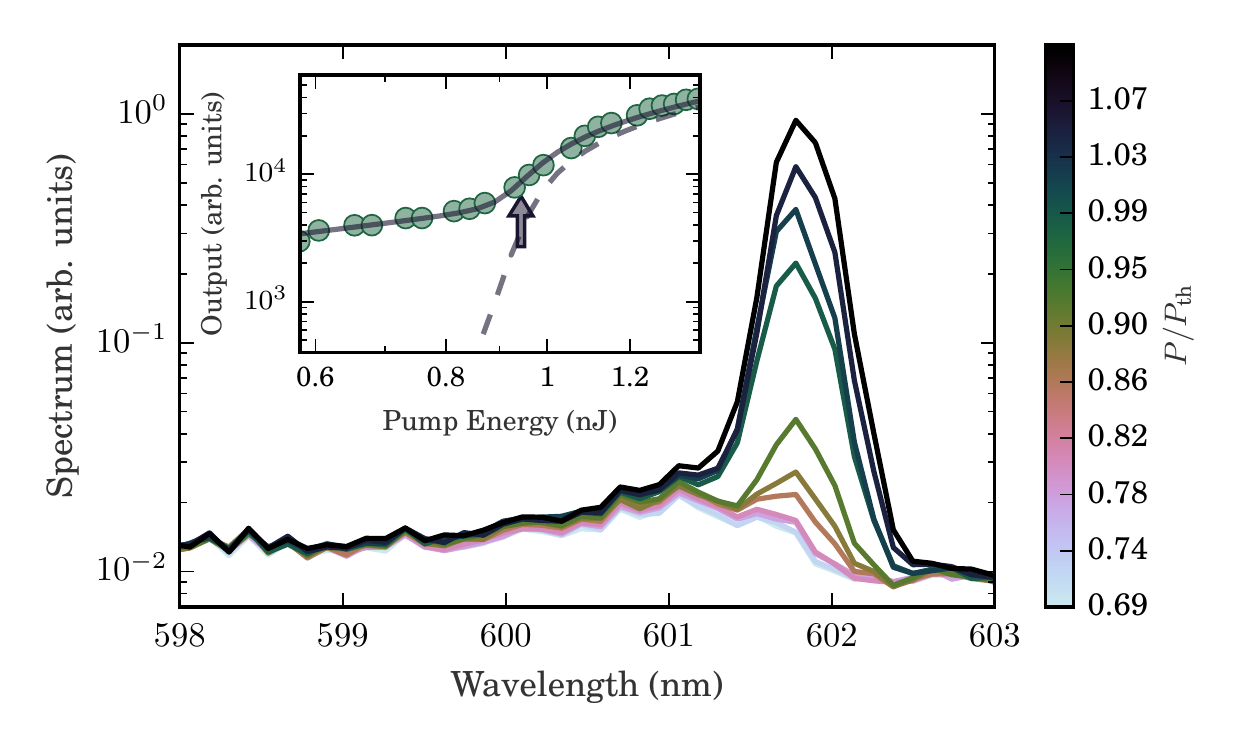}
\caption{Spectrum of the cavity output, above and below the condensation threshold at the critical excitation energy $P_\text{th}$. The spectral peak is located at the fundamental mode (groud-state) of the cavity, or cavity cutoff, at approximately 602~nm. The inset depicts the total cavity output (dots) and comparison with the single-mode mean-field model, with (full line) and without (dashed line) the contribution from the spontaneous emission background. The arrow indicates the threshold point.}
\label{in_out}
\end{figure}
\begin{figure*}
\centering
\includegraphics[width=0.8\textwidth]{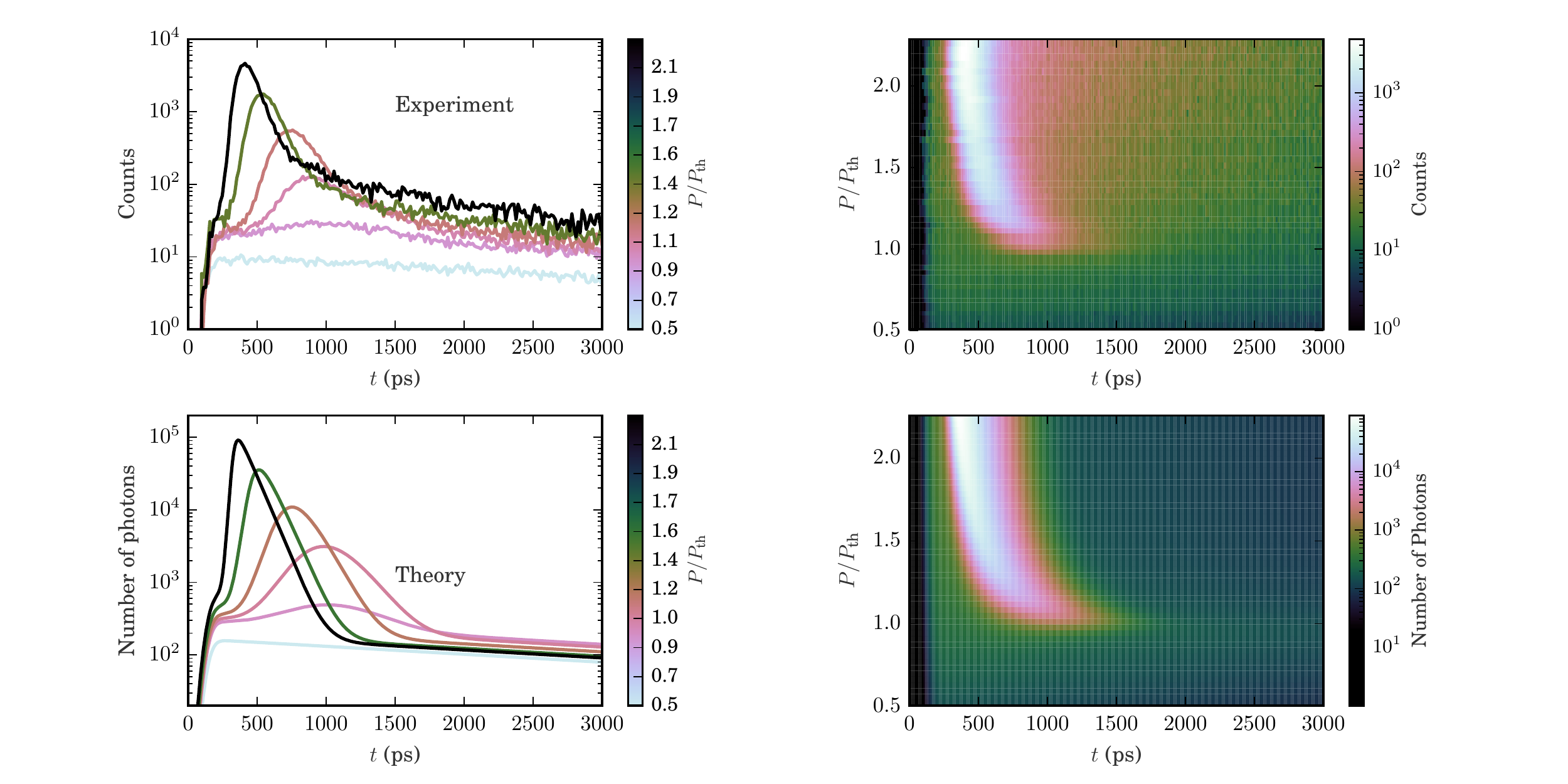}
\caption{Output light intensity as a function of the time following the pump pulse. We observe a delay in the growth of the photon condensate close to the critical excitation fraction $P_\text{th}$, accompanied by a large pulse broadening. Experimental results are shown on top and simulations of the mean-field rate equations on the bottom. The cavity ground state is located at 602 nm (cavity cutoff), the same conditions as in Fig.~(\ref{in_out}). The left panel depicts a subset of the data on the right.}
\label{mf_exp}
\end{figure*}
\par
We begin by demonstrating the existence of a (condensation) phase transition in the total amount of light emitted by the cavity as the excitation energy, or pump energy, $P$ is increased beyond a critical value $P_\text{th}$, as shown in Fig.~(\ref{in_out}). The spectra also show a tendency towards thermalization, witnessed by robust condensation in the cavity ground-state and indicating a regime where photon reabsorption plays a significant role~\cite{schmitt2015thermalization, Hesten2018}. Consequently, and despite the absence of a full thermal distribution, parallels may be drawn with Bose-Einstein condensation of photons~\cite{klaers2010, marelic2015, marelic2016, Greveling2018}. Despite only being strictly defined in thermal equilibrium as the macroscopic occupation of the ground state, we are assuming here a broader concept of condensation, as discussed in such diverse fields like physics, ecology, network theory or social sciences~\cite{Krapivsky2000, Chowdhury2000, Knebel2015, marelic2016}. This can be thought as the process where a particular, or small set of modes, in a multi-mode system becomes macroscopically occupied while the remaining ones saturate or become depleted. 
\par
By counting the rate of detection events in B1 and B2, we measure the total cavity output as a function of input pulse energy (input-output, or light-yield, curves), as shown in the inset of Fig.~(\ref{in_out}). The threshold, or critical, excitation energy can then be defined as the inflection in the light-yield curve. The signal from the cavity is collected without any filtering and multimode fibres are used to couple light into the detectors. Since non-condensed modes are also coupled to the APDs, we model their contribution by defining the detected signal $I_\text{det}$ as the sum of the single condensing cavity mode and a background of spontaneous emission, $I_\text{det} \propto \kappa n + \alpha \Gamma_\downarrow f N_\text{mol}$. Here, $\alpha$ is an empirical parameter set by fitting the light-yield curve to the mean-field rate equations. Loosely speaking, $\alpha$ determines the fraction of spontaneous emission into non-modelled cavity modes and is expected to be a small contribution, which will be verified in Sec.~(\ref{ensemble_av}).
%
%
\subsection{One-time Statistics}
\label{ensemble_av}
\begin{figure}
\centering
\includegraphics[width=0.45\textwidth]{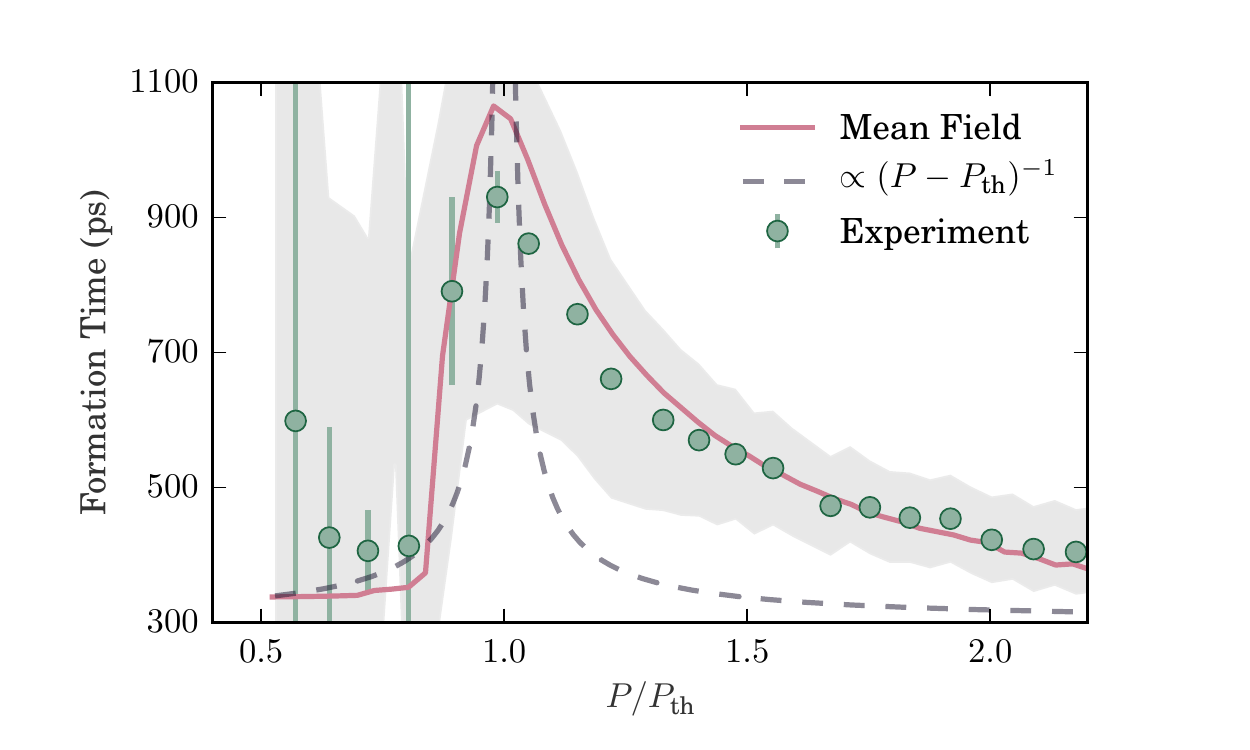}
\caption{Condensate formation time, defined as the interval between the pump pulse and the peak photon number, displaying a transient analogue of critical slowing down. The formation time and corresponding error bars are obtained by fitting Gaussian profiles in a neighbouring region around the peak photon number. The full-red and black-dashed lines are the mean-field simulations and the critical (power-law) divergence of condensate formation time for the lossless case, respectively. The shaded area depicts the increasing pulse width upon approaching the critical excitation energy.}
\label{critical_slowing}
\end{figure}
\begin{figure*}
\centering
\includegraphics[width=0.8\textwidth]{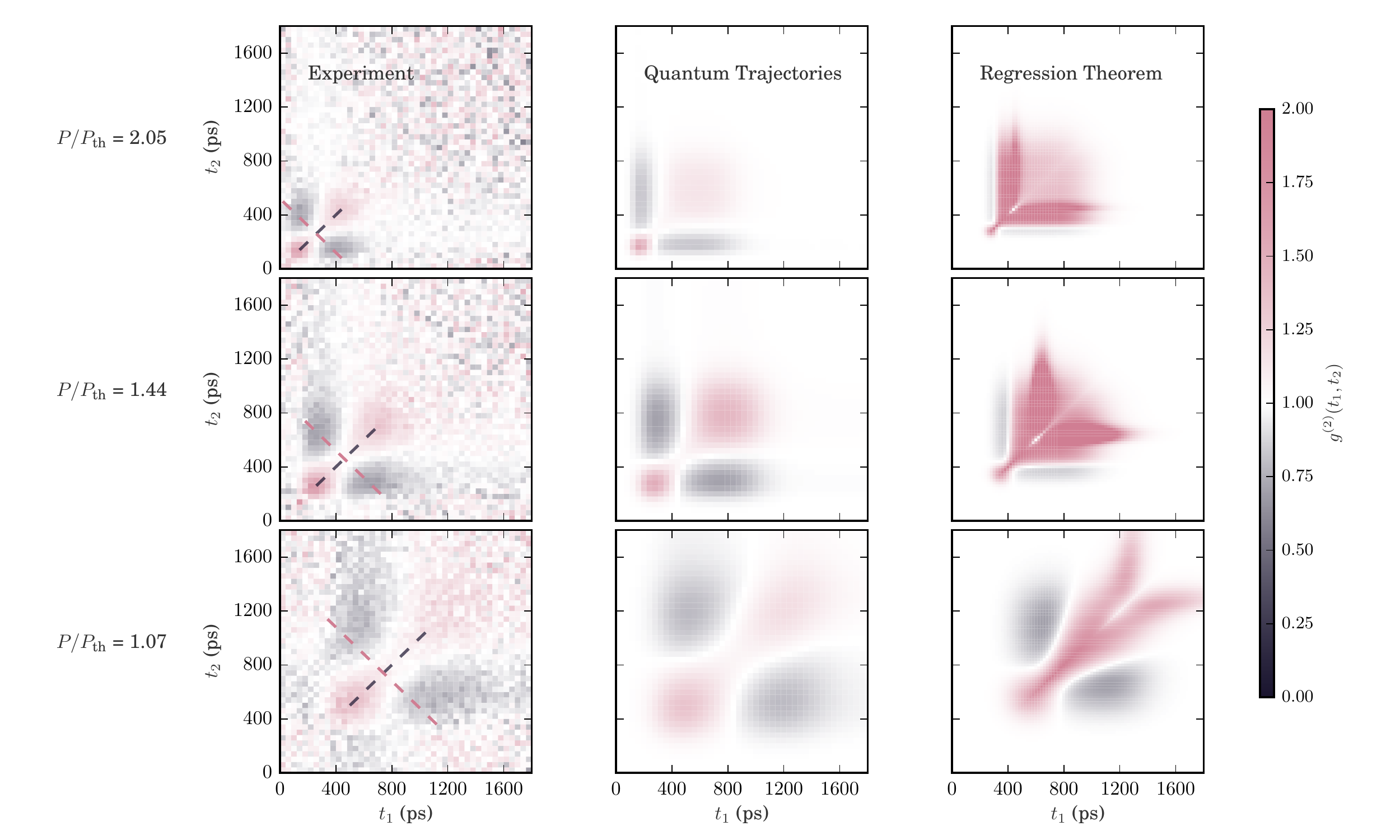}
\caption{Two-time, non-stationary, second-order correlation function, $\gTwo(t_1, t_2)$, for various excitation energies. Experimental data (left column) can be compared with quantum trajectories simulations (middle column) and a semi-analytic quantum regression approach based on a master equation expansion up to second-order (right column). In both cases, good agreement with the experiment requires a spontaneous emission background, as discussed in Sec.~(\ref{threshold}). The dashed curves in the experimental data correspond to the diagonal and off-diagonal regions of $\gTwo(t_1, t_2)$ depicted in greater detail in Fig.~(\ref{g2_cuts}). The cavity cutoff is set to $\lambda_0=$ 595 nm. Despite the slightly lower cutoff wavelength than that of Sec.~(\ref{ensemble_av}), the qualitative features of the average cavity output are the same as before.}
\label{g2_plot}
\end{figure*}
\par
Here, we expand the time-averaged results of the previous section to the time-dependent cavity output pulse shape, as shown in Fig.~(\ref{mf_exp}). By collecting unlabelled detections on both B1 and B2, we effectively average over any form of correlations and fluctuations. Pulses that form below the threshold excitation energy ($P<P_\text{th}$) display a simple exponential decay at a time scale of about $\tau_0\sim$~4~ns, the molecular excited-state lifetime. Above threshold, stimulated emission becomes important, leading to a large increase in photon number, followed by rapid depopulation of the condensate before a final decay at the slower time scale of the molecular excited state decay.
\par
It is instructive at this point to reflect about the interplay and coupled dynamics of molecular excitation fraction, $f$, and the number of cavity photons, $n$. In equilibrium, the molecular excitation fraction, $f$, cannot exceed its critical value, $f_c$. Under non-equilibrium conditions, however, if at any instant $f > f_c$ (e.g. after a quench), the photon population will grow exponentially until $f$ drops below $f_c$. The specifics of this relaxation into equilibrium, dictated at a mean-field level by Eqs.~(\ref{ndot}) and (\ref{fdot}), can show qualitatively different behaviour depending on the parameters of the cavity. First, we consider the case where $\Gamma_{\downarrow}$ is large compared to both $\kappa$ and $E$. The evolution of the molecular excitation fraction is dominated by the first term in Eq.~(\ref{fdot}) and the molecules act as a Markovian bath of excitations, in the sense that their dynamics are independent of $n$. The cavity population continues to grow until loss through the $\Gamma_{\downarrow}$ term causes $f$ to decrease below $f_c$. On the other hand, when $\Gamma_{\downarrow}$ is comparable or small in relation to $E$ and $\kappa$, and the system is quenched to a state with $f>f_c$, emission into the cavity plays a significant role in $f$ decreasing below $f_c$ and the molecules act as a non-Markovian excitation bath, retaining memory about the lost photons. This gives rise to non-trivial correlations in the cavity output, as will be demonstrated in Sec.~(\ref{g2_exp}).
\begin{figure*}
\centering
\includegraphics[width=0.80\textwidth]{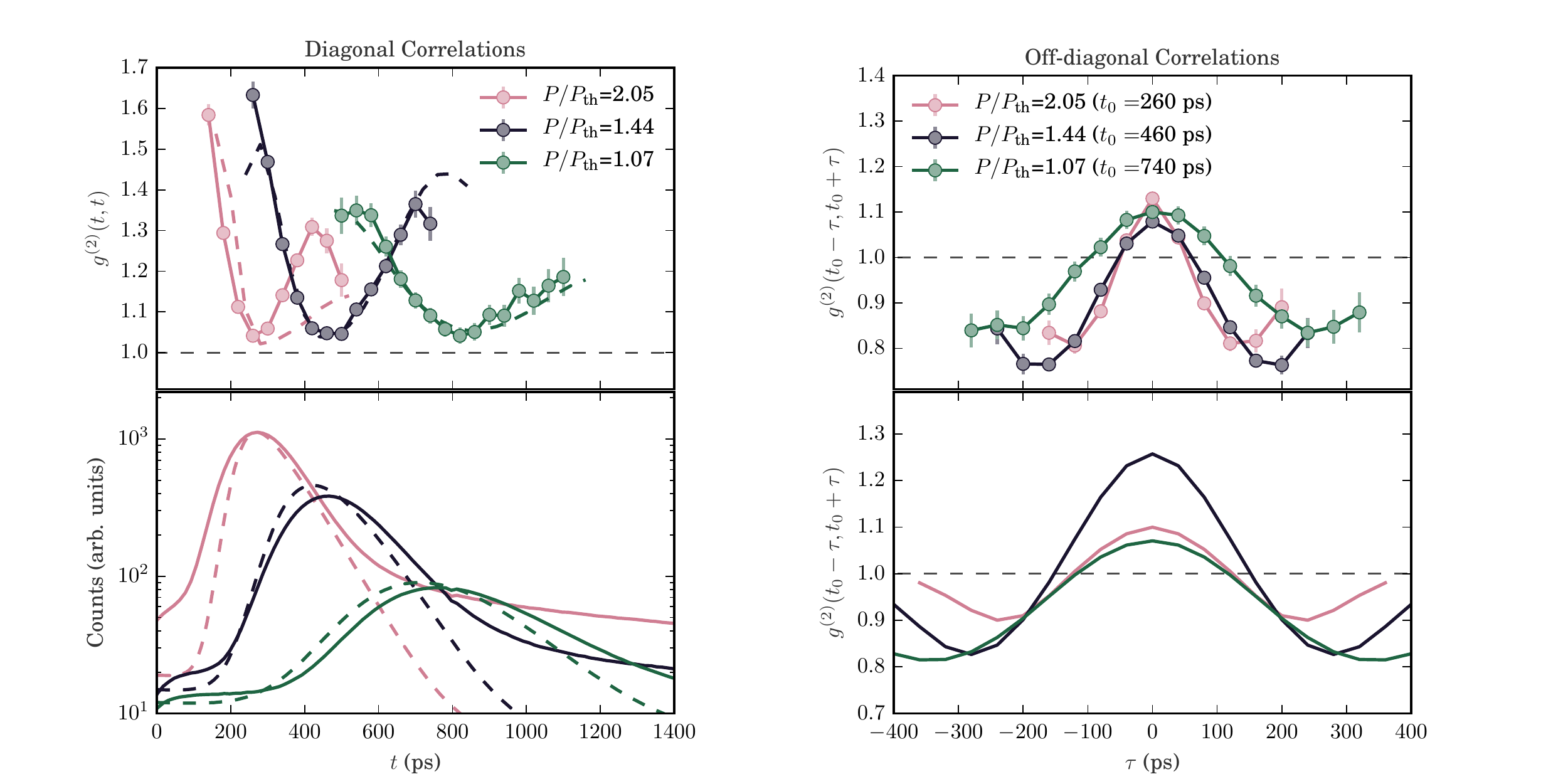}
\caption{Diagonal and off-diagonal correlations in the regions depicted by the dashed lines in Fig.~(\ref{g2_plot}). Top left: diagonal correlations, $g^{(2)}(t, t)$, shown on top, and the average condensate pulse, shown on the bottom left. The full and dashed lines depict the experimental data and the quantum trajectories results, respectively. Top and bottom right: off-diagonal correlations, $g^{(2)}(t_0-\tau, t_0+\tau)$, with $t_0$ the peak time of the average condensate pulse. Top right depicts the experimental results, with the quantum trajectories simulation on the bottom right. Here, the theory slightly deviate from the experimental results, which may be attributed to the error that propagates from determining the peak time $t_0$.}
\label{g2_cuts}
\end{figure*}
\par
We fit $\Gamma_{\downarrow}$ and $\kappa$ to the results in Fig.~(\ref{mf_exp}) and find the non-Markovian regime described earlier, with further evidence provided in Sec.~(\ref{g2_exp}). In particular, $\kappa = $10$^{10}$ s$^{-1}$, corresponding to a cavity lifetime of 100~ps, and $\Gamma_{\downarrow} = $0.998$ \Gamma_0$, with $\Gamma_0 = 1/\tau_0$ the molecular fluorescence decay rate, meaning that only 0.2 percent of the total molecular emission couples to the cavity. Together with the light-yield curves in Fig.~(\ref{in_out}), the contribution of spontaneous emission coupling into the detectors is found to be $\alpha=0.13$; small as expected. The emission and absorption rates are not taken as fitting parameters but rather calculated from experimental absorption and emission data for Rhodamine-6G~\cite{zenodo}. The total number of molecules is calculated from the dye concentration and cavity volume to be $\nmol =$1.9$\times$10$^8$.
\par
As $f$ approaches $f_c$ from above, the cavity dynamics become slow, as dictated by Eq.~(\ref{ndot}). In particular, since $\dot{n} \propto f - f_c$, one might expect \textit{critical slowing down} in the condensate formation time, with a critical exponent of -1, assuming $\dot{f} \sim 0$~\cite{slowPRL}. However, the presence of direct spontaneous emission into free space prevents $f$ from remaining close to $f_c$ for long times and the entire relaxation process is necessarily transient. Despite the mechanism of true \textit{critical slowing down} being frustrated, we still observe a slowing in the time taken for the condensate to form as we approach the critical excitation energy from above, as shown in Fig.~(\ref{critical_slowing}). By comparing to the critical divergence for the lossless case, where formation time is proportional to $(P-P_\text{th})^{-1}$, we observe a broadening of the threshold region due to the lossy nature of the cavity. Besides this transient analogue of critical slowing down, a distinct feature emerges upon approaching the critical excitation energy, the broadening of the average output pulse. In the next section, we show that this originates from a particular form of fluctuations that arise in such transient phase transitions: jitter in the condensate formation time.
%
%
%
\subsection{Two-time Statistics}
\label{g2_exp}
\par
Correlations and fluctuations of the cavity output can now be investigated by retaining the labelling of detection timestamps in B1 and B2. We then construct the two-time, non-stationary, second-order correlation function \gTwo($t_1, t_2$). Second-order correlations are typically described by the single-time \gTwo($\tau$) function, with $\tau = t_1 - t_2$, due to time-translation symmetry in steady-state conditions. In transient systems, however, the absence of this symmetry means that the full two-time, $t_1$ and $t_2$, dependence must be retained. We can then define
\begin{equation}\label{eq:g2}
\gTwo(t_1, t_2) = \frac{\langle a^{\dagger}(t_1)a^{\dagger}(t_2)a(t_2)a(t_1)\rangle}{\langle a^{\dagger}(t_1)a(t_1)\rangle\langle a^{\dagger}(t_2)a(t_2)\rangle}\approx \frac{P(t_1, t_2)}{P(t_1)P(t_2)}, 
\end{equation}
where $P(t_1, t_2)$ is the joint probability of photon detection at times $t_1$ and $t_2$ in detectors B1 and B2, respectively. By marginalizing over the second detector, $P(t_1)$ and $P(t_2)$ are obtained as the single-detector probabilities. The approximation in Eq.~(\ref{eq:g2}) is accurate as long as $[a^{\dagger}(t_1), a(t_2)] \approx 0$ or $\langle a^{\dagger}(t)a(t)\rangle \gg 1$. The former is satisfied when $|t_1 - t_2|$ is larger than the coherence time (much smaller than all relevant time scales involved in the cavity dynamics), and the latter is true for large photon numbers, as verified in Fig.~(\ref{mf_exp}).
\par
The second-order correlation function is shown in Fig.~(\ref{g2_plot}). The two main features to be retained here are the diagonal positive correlation ($g^{(2)}>1$) and the off-diagonal anti-correlation ($g^{(2)}<1$) lobes. These features are mainly a manifestation of the same kind of fluctuations -- \textit{jitter}, or shot-to-shot timing fluctuations, in the condensate formation -- which become amplified near the critical excitation energy. In the remainder of this section, we discuss this effect associated with transient phase transitions.
\par
Let's proceed by separately analysing diagonal and anti-diagonal correlations, as shown in Fig.~(\ref{g2_cuts}). For equal times, $g^{(2)}$ provides immediate information on number, or intensity, fluctuations, namely $g^{(2)}(t,t) \simeq 1 + \langle \Delta n(t) ^2 \rangle / \langle n(t) \rangle^2$. As such, periods of larger fluctuations coincide with the inflection point of the average pulse shape, consistent with a condensate forming at slightly different instants in each realization of the experiment. In a microscopic picture of the cavity dynamics, spontaneously emitted photons are required to seed the condensate growth. The randomness associated with the quantum nature of spontaneous emission then leads to such shot-to-shot time fluctuations, or jitter in the condensate formation. As we shall demonstrate in the next section, these periods of larger fluctuations correspond to a passage through the convex part of an effective free-energy landscape.
\begin{figure}
\centering
\includegraphics[width=0.5\textwidth]{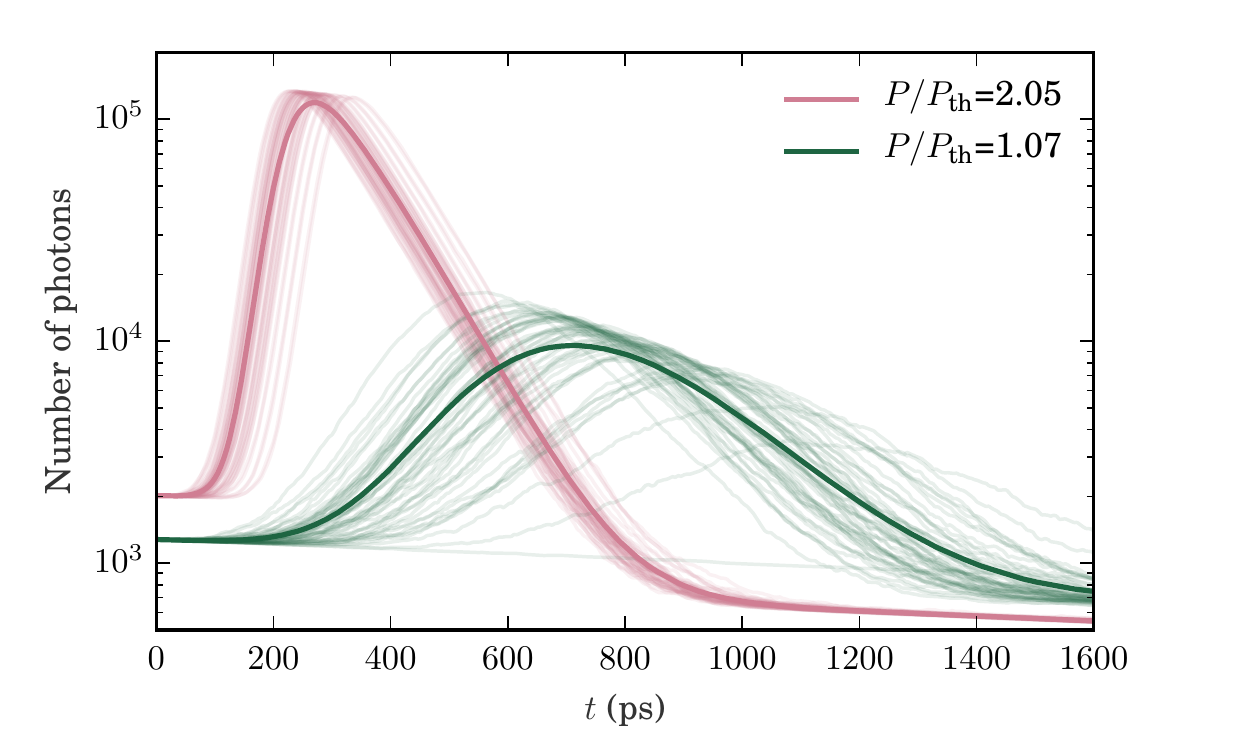}
\caption{Quantum trajectories simulation. 50 Individual trajectories are shown in light colors while darker colors depict their (ensemble) average. The parameters match those of Figs.~(\ref{g2_plot}) and (\ref{g2_cuts}).}
\label{montecarlo}
\end{figure}
\par
The above interpretation is further supported by the off-diagonal anti-correlation lobes, as seen in Figs.~(\ref{g2_plot}) and (\ref{g2_cuts}). Given the finite duration of the condensate pulse, if a photon is detected at an early time, it is less likely that another photon will be detected at a later time. In other words, the whole light pulse is either early or late. Off-diagonal regions with $g^{(2)} < 1$ are then an immediate witness of fluctuations in formation time. Also, it is further evidence that the bath of molecular excitations retains information about the history of the photon number, and is therefore non-Markovian, as discussed previously.
\par
By retaining correlations up to second-order in Eq.~(\ref{M_eq}) using a cluster expansion (or higher-order cumulants) \cite{Fricke1996, Gies2007, Kubo1962, Zens2019}, the two-time correlation function can be obtained via the quantum regression theorem~\cite{Gardiner2004}, as shown in Fig.~(\ref{g2_plot}). Details of this approach can be found in Appendix~(\ref{app2}). An alternative method to capture correlations to all orders is to construct a quantum trajectories (or Monte Carlo wavefunctions) approach~\cite{Carmichael1993, Dalibard1992,Dum1992,Molmer1993}. Different classes of events (molecular emission and absorption, cavity loss, etc) are defined and drawn at random given their respective rates. These are dynamically calculated as the number of photons and molecular excitations are updated at each step. Full details of this model can be found in Appendix~\ref{app3}. In the limit of a large number of realizations, this approach is equivalent to evolving the density matrix according to Eq.~(\ref{M_eq}). By retaining correlations to all orders, the quantum trajectories show the best match to data. In particular, while the second-order approximation does qualitatively predict the existence of correlation and anti-correlation regions, the results indicate that higher-order correlations quantitatively influence the cavity dynamics. However, the quantum regression approach, given its semi-analytical character, allows a faster sampling of different parameters.
\par
The experiment does not allow direct access to individual trajectories, only the effect of their relative fluctuations on $g^{(2)}$. However, the quantum trajectories method allows us to easily appreciate the effect of formation jitter, depicted in Fig.~(\ref{montecarlo}). The good agreement between experiment and theory in Figs.~(\ref{g2_plot}) and (\ref{g2_cuts}) is evidence that individual trajectories in the experiment have a similar form to those depicted in Fig.~(\ref{montecarlo}). Here, the formation jitter becomes clear, with larger shot-to-shot fluctuations occurring close to the critical excitation energy, which is at the origin of the pulse broadening described in Sec.~(\ref{ensemble_av}). The exact form of $g^{(2)}(t_1, t_2)$ depends on both the individual pulse shapes and their uncertainty in formation time. As it turns out, the earlier forming pulses (relatively far above threshold) are of shorter duration than later forming pulses (close to the critical point). This effect competes with the larger fluctuations in formation time closer to threshold, such that a diverging behaviour may not be extractable from the $g^{(2)}$ maps alone.
%
%
%
%
\section{Effective Free-Energy}
\label{FE}
\begin{figure}[b]
\centering
\includegraphics[width=0.45\textwidth]{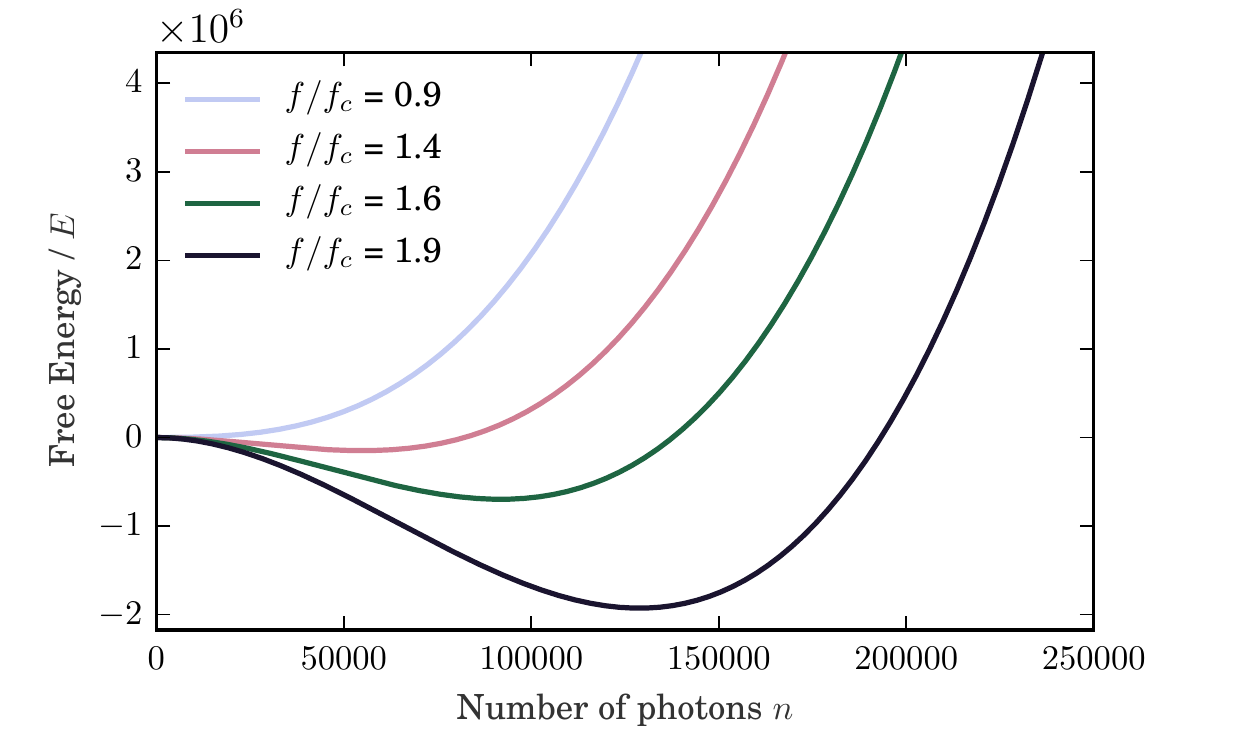}
\caption{Effective free-energy landscape for the microcavity photon number (order parameter), calculated for different values of the initial molecular excitation fraction $f$. Besides being the quantity directly varied in the experiment, $f$ determines the control parameter as $N_\text{ex} = n + f N_\text{mol}$}
\label{FE_plot}
\end{figure}
\par
The results presented in Sec.~(\ref{experiment}) may now be re-interpreted in a more general thermodynamic framework. In that sense, a free energy may be constructed as a function of an order parameter, the geometry of the former determining the evolution of the latter. While free energies are only strictly defined in thermodynamic equilibrium, there are effective analogues for non-equilibrium systems~\cite{allahverdyan2017free, degiorgio1970, agarwal1982higher}. In particular, the defining property of an effective free-energy $F$ is that it determines the average dynamics of the order parameter $\psi$ as~\cite{degiorgio1970}
\begin{equation}\label{eq:fe1}
\frac{d \psi}{dt} = - \frac{\partial \text{F}}{\partial \psi}.
\end{equation}
If a particular microscopic model is known, in the form of rate equations, for instance, one can reverse engineer Eq.~(\ref{eq:fe1}) to define an effective free-energy landscape~\cite{degiorgio1970}. Besides the formal equivalence between the two descriptions, the free-energy encodes all the relevant dynamical features in its geometrical properties while, at the same time, becoming independent of the microscopic details of a particular system. This allows us to predict universal properties for systems sharing similar free-energy geometries. Also, fluctuations can be built back into this simple model by including a stochastic term $\eta(t)$ in Eq.~(\ref{eq:fe1}) and turning it into a Langevin equation. 
\begin{figure}
\centering
\includegraphics[width=0.45\textwidth]{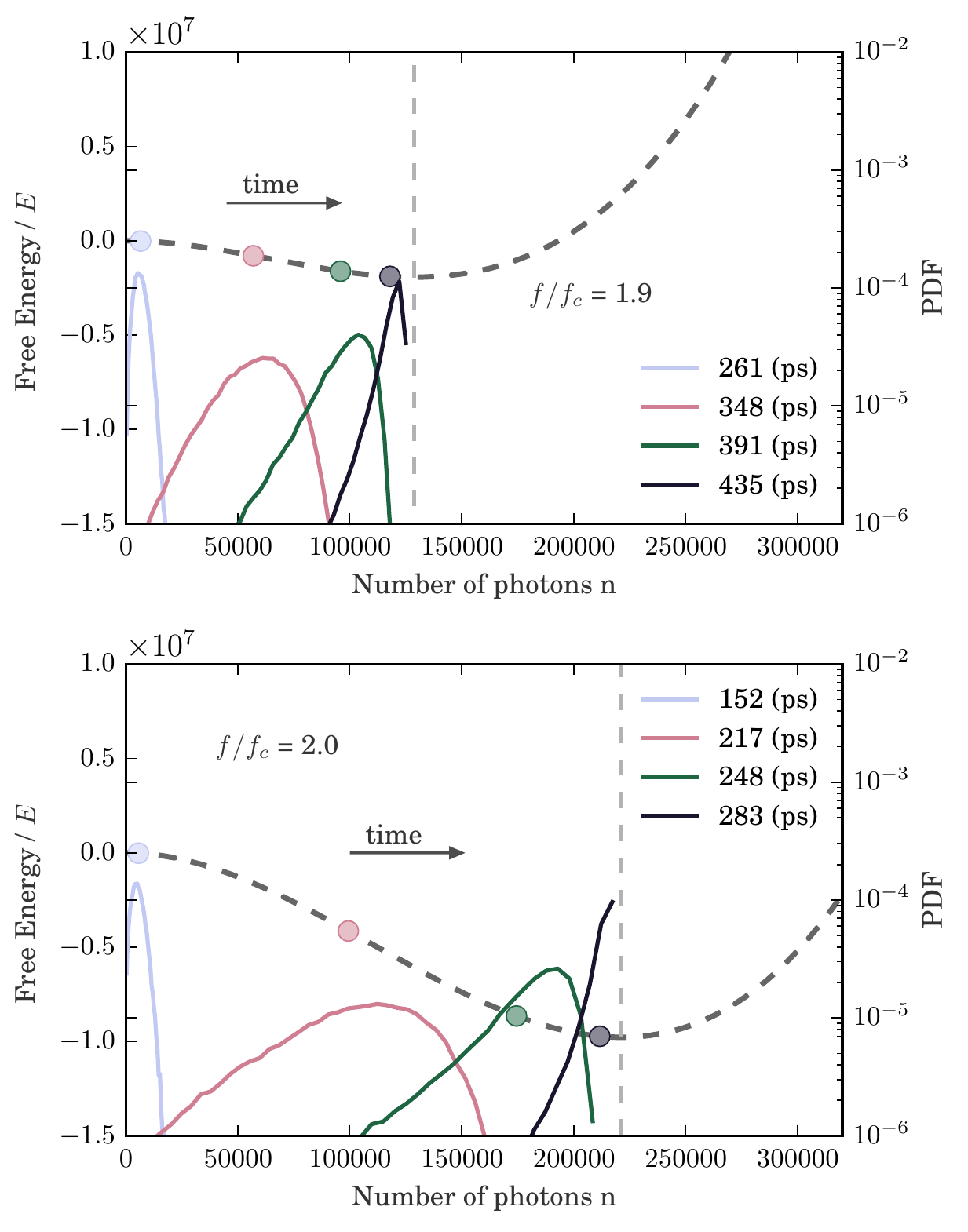}
\caption{Probability distribution function (PDF) of photon number (colored lines), describing fluctuations of the order parameter. The circles indicate the expected value of $n$ at a given time instant. The random walks (50000) are initialized at $n=0$ and the descent through the effective free-energy landscape (dark lines), as defined in Eq.~(\ref{eq:fe2}), is seeded by spontaneous emission. While the latter is essentially Gaussian noise, the free-energy convexity far from the equilibrium point induces non-Gaussian, heavy-tailed statistics in photon number. This skewed statistics partially explains the limitations of the quantum regression model described in Sec.~(\ref{experiment}).}
\label{pdf_evo}
\end{figure}
\par
Going back to Eqs.~(\ref{ndot}) and (\ref{fdot}), and ignoring cavity losses and pumping ($\kappa = \Gamma_{\downarrow} = \Gamma_{\uparrow} \equiv 0$), we find the effective free-energy for the non-equilibrium cavity dynamics as
\begin{eqnarray}\label{eq:fe2}
\text{F}(n) &=&  - \int_0^{n} \dot{n'}~dn' = \nonumber \\ 
&-& \frac{E \nex}{\nmol} n - \left[ \frac{E (\nex - 1)}{\nmol} - A \left(1 - \frac{\nex}{\nmol}\right) \right] \frac{n^2}{2} \nonumber \\
&+& \frac{(E + A)}{\nmol}\frac{n^3}{3}.
\end{eqnarray}
Here, the order parameter is the number of cavity photons ($\psi \equiv n$), while the control parameter is given by the total number of cavity excitations, as defined earlier. Fig.~(\ref{FE_plot}) depicts the free-energy for different (initial) excitation fractions, which allows for a direct analogy with the experiment, where the pump energy determines the initial excitation fraction, with the cavity being initialized with $n=0$. Moreover, all the parameters used here match those of Sec.~(\ref{experiment}).
\par
We consider the evolution of a probability distribution function (PDF) across the free-energy surface defined by Eq.~(\ref{eq:fe2}). This PDF encodes qualitative information about the fluctuation properties of the order parameter as it transiently evolves from the far-from-equilibrium state that follows a quench. We consider a large set of random walks evolving over the free-energy landscape. These walks (in photon number space) obey the Langevin equation described before where $\eta(t)$ accounts for spontaneous emission, such that $\langle \eta(t) \eta(0) \rangle = f^2 E^2 \delta (t)$~\cite{Druten2000, Mork2018}. From this ensemble of trajectories we estimate the order parameter PDF, as shown in Fig.~(\ref{pdf_evo}). An alternative, and formally equivalent, approach would be the construction a Fokker-Planck equation with a drift term given by the derivative of the free energy, as in Eq.~(\ref{eq:fe1}), and a diffusion term describing the fluctuations from spontaneous emission.
\par
Quite generally, fluctuations act to broaden the photon number PDF while a positive (concave) curvature in the free-energy tends to localise it. At a second-order phase transition, the curvature at the minimum of the free energy (defining an equilibrium order parameter) vanishes and the PDF shows diverging fluctuations which persist for long times, giving rise to critical slowing down. In transient, non-equilibrium systems dramatic features also occur, with regions of negative (convex) curvature acting to amplify fluctuations. A PDF evolving through these regions while relaxing towards the free-energy minimum experiences a short-lived but large increase of fluctuations, as shown in Fig.~(\ref{pdf_evo}). The jitter described in Sec.~(\ref{experiment}) is the immediate consequence of this. The maximum of number fluctuations, marked by the peak in $g^{(2)}(t,t)$, and shown in Fig.~(\ref{g2_cuts}), occurs when the average order parameter reaches the free-energy inflection point. From Eq.~(\ref{eq:fe1}), this corresponds to the (temporal) inflection point of the order parameter, in complete agreement with the results depicted in the previous section. The width of the photon number PDF then shrinks back as the order parameter evolves to the concave part of $F$. While the free-energy landscape is more convex for larger values of the control parameter, which increases number fluctuations, these are longer-lived close to threshold. At the level of individual realizations, this means that the timing jitter in the formation of the condensed phase is larger close to the critical excitation energy.
\par
In the presence of loss by cavity transmission, the free-energy and photon number PDF are coupled in a nontrivial way. For sufficiently large $\kappa$, as in the case of the experiment described in Sec.~(\ref{experiment}), the rate of change of the free energy landscape depends on the photon number $n$, which is itself described by a given PDF. The landscape is now neither constant, nor a simple function of time, but rather coupled to the photon number history, such that for trajectories where the condensate forms early, it also decays early, leading to the anti-correlation lobes seen in $\gTwo(t_1, t_2)$. This is essentially the same result as depicted by the quantum trajectories simulation in Fig.~(\ref{montecarlo}) but reinterpreted under the geometrical properties of the effective free-energy landscape.
\par
The free-energy description assumes the total number of cavity excitations $N_\text{ex}$, the control parameter, to be fixed, such that all molecular excitations are converted into cavity photons. In the experiment, this approximation is only valid in the non-Markovian regime described earlier, where molecular de-excitation by stimulated emission dominates over direct emission into free-space. In the Markovian regime, where losses of molecular excitation are dominated by the $\Gamma_\downarrow$ term, both critical slowing down and timing jitter in the condensate formation vanish. As a final remark, Eq.~(\ref{eq:fe1}) allows for a formal reconstruction of the free-energy landscape, $F(\psi)$, from the observed average dynamics of the order parameter, $\psi(t)$, although in practice the results are not very informative.
%
%
\section{Conclusions}
\label{conclusions}
\par
In this work, we have described the transient non-equilibrium dynamics of light in a dye-filled optical cavity quenched through a condensation phase transition. By rapidly exciting a large number of dye molecules, the system is brought to a far-from-equilibrium state. By averaging over all forms of fluctuations, we observed a delayed formation of the condensed phase, interpreted as a transient equivalent of critical slowing down. When quenched above the condensation threshold excitation energy, the quantum fluctuations associated with spontaneous emission seed the growth of the order parameter as the system relaxes into equilibrium. The relaxation dynamics is slower close to the critical point, a feature easily interpreted under the geometrical properties of the effective free-energy landscape, which becomes flat. The same mechanism is responsible for the usual critical slowing down in the relaxation rate of the ordered phase that follows a second-order phase transition. Also, despite the absence of latent heat and the fact that we are dealing with second-order and not first-order phase transitions, analogies can be drawn with the precipitation in supercooled, or supersaturated, liquids. Even quenched above the critical point, a seed of spontaneously emitted photons is needed to nucleate condensation, playing the role the seeding crystals in supercooled, or supersaturated, liquids. Also, once seeded, crystallization across the entire liquid is faster for liquids quenched further across their critical parameters, with temperature playing the same role as the excitation fraction that follows the quench, in the optical cavity context.
\par
By measuring the statistical properties of this transient condensation, we describe a novel form of diverging fluctuations around the critical point, jitter in the formation of the ordered phase. These are witnessed by strong diagonal correlations and off-diagonal anti-correlations in the non-stationary, second-order correlation function, $g^{(2)}(t_1, t_2)$. More precisely, we demonstrated that while the diagonal of $g^{(2)}$ is a powerful probe of the geometrical properties of the free-energy landscape, its off-diagonal elements reflect the relevant dissipation processes, with the anti-correlation lobes a joint effect of jitter and cavity loss. Fluctuations, arising from spontaneous emission, are highly amplified as the order parameter goes through the convex part of the free-energy landscape towards its equilibrium point. The condensation jitter is a direct physical manifestation of this increasing fluctuations, in accordance with the ideas put forward in  Ref.~\cite{jaynes1986predictive}. 
\par
The description in terms of the geometric properties of the effective free-energy landscape, being independent of the microscopical details of our particular system, allows us to generalize our observations. In particular, both the transient critical slowing down and the jitter in the formation of the order parameter are expected to be universal features of the dynamics that follows a quench through a second-order phase transition. In micro- and nano-lasers, in particular, the full two-time, non-stationary analysis of the relaxation process has been greatly overlooked and previous results~\cite{Ulrich2007, Assmann2009, Wiersig2009, Assmann2010} may now benefit from being re-examined. Despite some recent efforts in this direction~\cite{Lebreton2013, Lebreton2015, Moody2018}, and to the best of our knowledge, we present here for the first time a generic and comprehensive description of the relation between temporal and number fluctuations in the non-stationary dynamics of systems undergoing second-order phase transitions. Implications of our results may also be of concern in the growth of colloidal nanoparticles. These result from a nucleation process in a supersaturated chemical solution, and the growth mechanism can be described by the minimization of a free-energy~\cite{Polte2015, Thanh2014}. Finally, the system studied in this work, as well as the related examples stated above can be described by single-value order parameters. One may wonder on the generalization of these effects in spatially extended systems, where the order parameter is a function of both space and time. 
%
%
\section*{Acknowledgements}
We acknowledge financial support from EPSRC (UK) through the grants EP/S000755/1, EP/J017027/1, the Centre for Doctoral Training in Controlled Quantum Dynamics EP/L016524/1 and the European Commission via the PhoQuS project (H2020-FETFLAG-2018-03) number 820392. We also thank Julian Schmitt for helpful discussions. The data related to this paper may be requested from the authors or via \href{mailto:dataenquiryEXSS@imperial.ac.uk}{\nolinkurl{dataenquiryEXSS@imperial.ac.uk}}.
%
%
\section{Appendix}
\appendix
\section{Second-order rate equations and the quantum regression theorem \label{app2}}
\par
From the non-equilibrium model introduced in Eq.~(\ref{M_eq}), one can derive rate equations for the ensemble-averaged photon number, $\langle n\rangle = \langle\hat{a}^\dag(t)\hat{a}(t)\rangle$, and the number of excited molecules, $\langle m\rangle = \sum_k \langle\sigma^+_k(t)\sigma^-_k(t)\rangle$, as
\begin{eqnarray}
\frac{d\langle n\rangle}{dt} &=& -\kappa \langle n\rangle + E\{\langle m\rangle + \langle nm\rangle\} - A\{\langle n\rangle \nmol -\langle nm\rangle\}\nonumber\\
\frac{d\langle m\rangle}{dt} &=&  - E\{\langle m\rangle + \langle nm\rangle\}+ A\{\langle n\rangle \nmol - \langle nm\rangle\} \nonumber\\
&-&\Gamma_{\downarrow} \langle m\rangle + \Gamma_\uparrow(\nmol -\langle m\rangle).
\label{rate_eq}
\end{eqnarray}
The calculation of $\langle nm\rangle$ depends on the estimation of $\langle n^2m\rangle$, $\langle nm^2\rangle$, $\langle n^3\rangle$, $\cdots$, which requires solving a large number of ordinary differential equations. These can be reduced with an hierarchical set of approximations. For instance, in the semi-classical limit, the expectation values for $n$ and $m$ are factorized, $\langle mn\rangle \approx \langle m\rangle\langle n\rangle$, reducing Eqs.~(\ref{rate_eq}) to
\begin{eqnarray}
\dot{n} &=& -\kappa n + E~f(n +1) - A~n (1-f) \nonumber\\
\dot{f} &=&  -\Gamma_{\downarrow}f  + A~n(1-f)- E~(n +1)f + \Gamma_\uparrow(1-f).\nonumber
\label{rate_eq_1}
\end{eqnarray}
These are equivalent to Eqs.~(\ref{ndot}) and (\ref{fdot}). Here, we define $f = \langle m\rangle/\nmol$ as the molecular excitation fraction and set, for the ease of notation, $n = \langle n\rangle$. Despite ignoring correlations all-together, this corresponds to a first level approximation to the non-equilibrium cavity dynamics. 
\par
In order to account for correlations and fluctuations, one needs to go beyond the semi-classical approximation. In particular, the expectation values can be expanded in a hierarchical manner~\cite{Fricke1996,Gies2007,Kubo1962,Zens2019} given by
\begin{eqnarray}
\sigma_{xy}^2 &=& \langle x y \rangle -  \langle x \rangle\langle y \rangle\\
\sigma_{xyz}^3 &=& \langle x y z \rangle - \sum\sigma_{xy}^2\langle z\rangle - \langle x\rangle\langle y\rangle \langle z\rangle \\
\sigma_{wxyz}^4 &=& \langle w x y z \rangle - \sum\sigma_{wxy}^3\langle z\rangle - \sum\sigma_{wx}^2 \langle y\rangle\langle z \rangle\nonumber\\
&-& \sum\sigma_{wx}^2\sigma_{yz}^2 - \langle w\rangle\langle x\rangle\langle y\rangle \langle z\rangle.
\end{eqnarray}    
These represent the second, third, and fourth order cumulants, with the summation referring to all possible combination of variables. A minimal description of correlations is constructed by truncating the hierarchy at second-order. In this way, and by defining $\sigma_{x}^2$ = $\langle x^2 \rangle -\langle x \rangle^2$, with $x = \lbrace n,m \rbrace $, we explicitly write
\begin{widetext}
\begin{eqnarray}
\dot{n} &=&-\kappa n + E\{(n+1)m +\sigma^2_{nm}\} - A\{n(\nmol -m)-\sigma^2_{nm}\}\label{second_1}\\ 
\dot{m}&=&-\Gamma_\downarrow m - E\{(n+1)m +\sigma^2_{nm}\}+
\Gamma_\uparrow(\nmol -m)
+A\{n(\nmol -m)-\sigma^2_{nm}\}\label{second_2}\\
\dot{\sigma}^2_{n}&=&-\kappa(n+2\sigma^2_n) + E\{(n+1)m +2\sigma^2_nm+\sigma^2_{nm}(2n+1)\} \nonumber\\
&-& A\{n(\nmol-m)+2\sigma^2_n(\nmol-m)-\sigma^2_{nm}(2n-1)\}\label{second_3}\\
\dot{\sigma}^2_{m}&=&-\Gamma_\downarrow(m+2\sigma^2_m) - E\{-(n+1)m +2\sigma^2_m(n+1)+\sigma^2_{nm}(2m-1)\} \nonumber\\
&+&A\{n(\nmol -m)-2\sigma^2_mn + \sigma^2_{nm}(-2m+2\nmol -1)\}+\Gamma_\uparrow(\nmol -m-2\sigma^2_m)\label{second_4}\\
\dot{\sigma}^2_{nm}&=& -(\kappa+\Gamma_\downarrow+\Gamma_\uparrow)\sigma^2_{nm} + E\{(n+1)(-m + \sigma^2_m)-\sigma^2_nm+\sigma^2_{nm}(m-n-2)\} \nonumber\\
&+& A\{-n(\nmol -m)+\sigma^2_mn+\sigma^2_n(\nmol -m)+\sigma^2_{nm}(m-n+1-\nmol )\}.\label{second_5}
\end{eqnarray}
\end{widetext}
The second-order photon correlation function at zero time delay, $g^{(2)}(t)$, follow immediately as
\begin{eqnarray}
g^{(2)}(t) &=& \frac{\langle\hat{a}^\dag(t)\hat{a}^\dag(t)\hat{a}(t)\hat{a}(t)\rangle}{\langle\hat{a}^\dag(t)\hat{a}(t)\rangle^2} = \frac{\langle n^2(t) \rangle-\langle n(t)\rangle}{\langle n(t)\rangle^2},\nonumber\\
&=& 1+\frac{\sigma_n^2(t)- n(t)}{n^2(t)}.
\label{g2}
\end{eqnarray}
\par
The two-time second-order correlation function can be obtained by invoking the quantum regression theorem~\cite{Gardiner2004}, which allows us to calculate any quantity of the form $\langle X(t+\tau) Y(t)\rangle$ using two single-time evolutions. Let the initial state of the system be $\chi(0)$, and the evolution be given by the map, $\chi(t)$ = $\mathcal{V}(t,t')\chi(t')$. The two-time expectation value can then be written as
\begin{eqnarray}
\langle X(t+\tau) Y(t)\rangle &=& \mathrm{Tr}[X\mathcal{V}(t+\tau,t)\{Y\chi(t)\}],  \nonumber\\
&=& \mathrm{Tr}[X\mathcal{V}(t+\tau,t)\{Y\mathcal{V}(t,0)\chi(0)\}].~~~~~~
\label{quant_reg}
\end{eqnarray}
The two-time function is thus calculated by evolving $\chi(0)$ from 0 to $t$, followed by the conditional state $Y\chi(t)$ from $t$ to $t+\tau$. For our cavity model, we begin by first evolving the density operator, $\rho$, from $t = 0$ to $t = t_1$, using the second-order rate Eqs.~(\ref{second_1}) and(\ref{second_5}), obtaining $g^{(2)}(t_1)$. Second, the first-order rate Eqs.~(\ref{ndot}) and (\ref{fdot}) are used to evolve the conditional state, $\tilde{\rho} = {\hat{a}(t_1) \rho \hat{a}^\dag(t_1)}/{\langle \hat{a}^\dag(t_1)\hat{a}(t_1)\rangle}$ from $t = t_1$ to $t=t_2$. Following Eq.~(\ref{quant_reg}), one then arrives at the two-time photon correlation function, 
\begin{equation}
g^{(2)}(t_1,t_2) =  \frac{\langle \hat{a}^\dag(t_1)\hat{a}^\dag(t_2)\hat{a}(t_2)\hat{a}(t_1)\rangle}{\langle \hat{a}^\dag(t_1)\hat{a}(t_1)\rangle ~\langle \hat{a}^\dag(t_2)\hat{a}(t_2)\rangle}.
\end{equation}
\section{Quantum trajectories approach \label{app3}}
\par
The second-order approach described above corresponds to a first-level approximation to the description of correlations and fluctuations in the cavity dynamics. Moving to higher-order expansions increases the number of ordinary differential equations needed to resolve the dynamics, which soon becomes cumbersome and impractical. An alternative approach to solve the master Eq.~(\ref{M_eq}), is to use the quantum trajectories (or Monte-Carlo wavefunction) method~\cite{Carmichael1993, Dalibard1992,Dum1992,Molmer1993}. Here, the Lindblad dynamics of the density operator $\rho$ is replaced by a wavefunction whose evolution  is given by a non-Hermitian effective Hamiltonian, interspersed with stochastic quantum jumps. Subsequently, evolution of $\rho$ is approximated by an ensemble-average of wavefunctions, or trajectories, say $ \lvert \psi_i\rangle$. For a large number of trajectories, $z$, the average of any observable is then given by
\begin{eqnarray}
\langle \hat{X}(t) \rangle = \mathrm{Tr}[\hat{X}\rho(t)] \approx \frac{1}{z}\sum_{i=1}^z \langle\psi_i(t)|\hat{X}|\psi_i(t)\rangle .
\end{eqnarray}
\par
The effective non-Hermitian Hamiltonian for the non-equilibrium cavity model in Eq.~(\ref{M_eq}) is given by
\begin{equation}
\mathcal{H}_\mathrm{eff} = H_0 - \frac{i}{2} \sum_k J_k^\dag J_k,
\label{Heff}
\end{equation}
where $J_k$ are the jump operators defining the stochastic dynamics. In the non-equilibrium cavity model, coherences cannot be created by $\mathcal{H}_\mathrm{eff}$. Hence, if a quantum trajectory starts in a particular number state, say $|\psi_i(0)\rangle = |n_0,m_0\rangle$, the action of $\mathcal{H}_\mathrm{eff}$ alone does not change the state in this number basis. The complete dynamics of the trajectory is simply governed by the stochastic jumps $J_k$, occuring at rates $R_k$:
\begin{eqnarray}
\sqrt{\kappa}~\hat{a}&:&|n,m\rangle \rightarrow |n-1,m\rangle; \quad R_0 = \kappa n\nonumber\\
\sqrt{\Gamma_\uparrow}~\sigma^+&:&|n,m\rangle \rightarrow |n,m+1\rangle; \quad R_1 = \Gamma_\uparrow (N-m)\nonumber\\
\sqrt{\Gamma_\downarrow}~\sigma^-&:&|n,m\rangle \rightarrow |n,m-1\rangle; \quad R_2 = \Gamma_\downarrow n\nonumber\\
\sqrt{E}~\hat{a}^\dag\sigma^-&:&|n,m\rangle \rightarrow |n+1,m-1\rangle; \quad R_3 = E (n+1)m \nonumber\\
\sqrt{A}~\hat{a}\sigma^+&:&|n,m\rangle\rightarrow|n-1,m+1\rangle; \quad R_4 = A n(N-m).\nonumber\\
\end{eqnarray}
A particular quantum trajectory is constructed by drawing a series of stochastic events, with their individual probabilities proportional to the rates $R_k$. The time between consecutive events is drawn from an exponential distribution, whose mean is the inverse of total rate of events. From a large ensemble of trajectories, we can calculate the non-stationary second-order correlation function, $g^{(2)}(t_1,t_2)$, as
\begin{equation}
g^{(2)}(t_1,t_2) = \frac{\langle n(t_1) n(t_2) \rangle}{\langle n(t_1)\rangle\langle n(t_2) \rangle},
\end{equation}
where $\langle \cdot \rangle$ denotes ensemble average, over the entire set of trajectories. The same approximations as discussed in Sec.~(\ref{g2_exp}) are assumed here as well.

%
\bibliographystyle{prsty}

%
\end{document}